\documentclass[aps,prd,twocolumn,showpacs]{revtex4}

\usepackage{graphicx}
\usepackage{dcolumn}
\usepackage{bm}
\usepackage{amsmath}
\usepackage{epstopdf}
\usepackage{amsfonts}
\usepackage{amssymb}
\usepackage{hyperref}
\usepackage{xcolor}
\usepackage{float}

\usepackage{natbib}

\providecommand{\U}[1]{\protect\rule{.1in}{.1in}}

\newcommand{\baa}{\begin{align}}
\newcommand{\eaa}{\end{align}}

\newcommand{\be}{\begin{equation}}
\newcommand{\ee}{\end{equation}}
\newcommand{\bea}{\begin{eqnarray}}
\newcommand{\eea}{\end{eqnarray}}

\begin{document}

\title{Greybody factors for a minimally coupled scalar field in three-dimensional Einstein-power-Maxwell black hole background}

\author{Grigoris Panotopoulos}
\affiliation{Centro de Astrof{\'i}sica e Gravita{\c c}{\~a}o, Instituto Superior T{\'e}cnico-IST,
Universidade de Lisboa-UL, Av. Rovisco Pais, 1049-001 Lisboa, Portugal}
\email{grigorios.panotopoulos@tecnico.ulisboa.pt}

\author{\'Angel Rinc\'on}
\affiliation{Instituto de F\'{i}sica, Pontificia Universidad Cat\'{o}lica de Chile, \mbox{Avenida Vicu\~na Mackenna 4860, Santiago, Chile.}}
\email{arrincon@uc.cl}

\date{\today}

\begin{abstract}
In the present work we study the propagation of a probe minimally coupled scalar field in Einstein-power-Maxwell charged black hole background in (1+2) dimensions. We find analytical expressions for the reflection coefficient as well as for the absorption cross-section in the low energy regime, and we show graphically their behaviour as functions of the frequency for several values of the free parameters of the theory.
\end{abstract}

\maketitle

\section{Introduction}

Hawking radiation \cite{hawking1,hawking2} is one of the most fascinating aspects of theoretical physics, since it is a manifestation of a quantum effect in curved spacetime, and as such it has always attracted a special interest in the community, although it has not been detected in the Universe yet. Black holes (BHs), a generic prediction of Einstein's General Relativity, are objects of paramount importance to gravitational theories, and excellent laboratories to study and understand various aspects of classical and quantum gravity.

Of particular interest is the so called greybody factor, or else the absorption cross section, which is a frequency dependent factor that
measures the modification of the original black body radiation, and thus gives us valuable information about the
near-horizon structure of black holes \cite{kanti1}. Consequently, in the literature exist many works in which the authors
have studied the propagation and relativistic scattering of probe fields in different gravitational backgrounds, and have analysed the
corresponding greybody factors, either in asymptotically flat spacetimes or in asymptotically non-flat spacetimes with a non-vanishing cosmological constant. For a partial list see e.g. \cite{col1,col2,col3,col4,col5,col6,col7,3D1,3D2,Fernando:2004ay,chinos,coupling,kanti2,kanti3,Panotopoulos:2016wuu,Panotopoulos:2017yoe,Ahmed:2016lou} 
and references therein.

Gravity in (1+2) dimensions is special and it has attracted a lot of attention
due to the absence of propagating degrees of freedom, and also due to its deep 
connection to a Yang-Mills theory with the Chern-Simons term only \cite{CS,Witten:1988hc,Witten:2007kt}. Since three-dimensional BHs have thermodynamic properties closely analogous to those of four-dimensional black holes \cite{review}, 
one may get some insight into realistic BHs studying a simpler system.

Non-linear electrodynamics in various contexts is quite interesting for several different reasons. For example, the Born-Infeld non-linear electrodynamics was originally introduced in the 30's in order to obtain a finite self-energy of point-like charges \cite{BI}. Interestingly enough, during  the  last  decades  this type of action reappears in the open sector of superstring theories \cite{ST1,ST2} as it describes the dynamics of D-branes \cite{Dbranes1,Dbranes2}. What is more, straightforward generalization of Maxwell’s theory leads to the so called Einstein-power-Maxwell (EpM) theory described by a Lagrangian density of the form
$\mathcal{L}(F) = F^k$, where $F$ is the Maxwell invariant, and $k$ is
an arbitrary rational number. Clearly the special value $k = 1$ corresponds to linear electrodynamics.

Currently, this class of non-linear electrodynamics has been receiving attention in several contexts \cite{Xu:2014uka,Mazharimousavi:2011nd,Rincon:2017goj,Panotopoulos:2017hns,Rincon:2018sgd}. The reason why studying such a class of theories is interesting lies on the fact that Maxwell's theory in higher dimensions is not conformally invariant, while in a D-dimensional spacetime the electromagnetic stress-energy tensor is traceless if the power $k$ is chosen to be $k=D/4$. Therefore in four dimensions the linear theory is conformally invariant, and this corresponds of course to the standard Maxwell's theory. In a three-dimensional spacetime, however, if $k=1$ the theory is linear but the electromagnetic stress-energy tensor is not traceless, whereas if $k=3/4$ the theory is conformally invariant but non-linear. Black hole solutions in (1+2)-dimensional and higher-dimensional EpM theories have been obtained in \cite{BH1} and \cite{BH2} respectively, and the greybody factor for the (1+2)-dimensional case with a non-vanishing cosmological constant was studied in \cite{chinos} for $k=2/3$.

In the present article we wish to analyse the propagation of a probe canonical massless scalar field into a three-dimensional gravitational spacetime, and obtain analytical expressions for the reflection coefficient and the corresponding greybody factor. Our work is organized as follows:  After this
Introduction, we present the theory and the corresponding black hole solution as well as the scalar wave equation in the next Section.   
In Section \ref{Sol} we obtain approximate analytical expressions for the reflection coefficient as well as for the greybody factor valid at low frequencies, and we also briefly discuss our numerical results. Finally, we conclude our work in Section \ref{Concl}.

\section{The background and the scalar wave equation}

\subsection{The theory and the black hole solutions}

We consider the theory in (1+2) dimensions described by the action
\begin{equation}
S[g_{\mu \nu}] = \int d^3x \sqrt{-g} \left[ \frac{1}{2 \kappa}R - (F_{\mu \nu} F^{\mu \nu})^k \right]
\end{equation}
where $\kappa \equiv 8 \pi G$ with $G$ being Newton's constant is the gravitational constant, $k$ is an arbitrary rational number, $R$ is the Ricci scalar, $g$ the determinant of the metric, and $F_{\mu \nu}$ the electromagnetic field strength. Varying the action with respect to the metric and the gauge field $A_\mu$ one obtains the field equations \cite{BH1,BH2}
\begin{eqnarray}
G_{\mu \nu} & = & 4 \kappa \alpha \left [k F_{\mu \rho} F_\nu ^\rho F^{k-1} - \frac{1}{4} g_{\mu \nu} F^k \right ] \\
0 & = & \partial_\mu (\sqrt{-g} F_{\mu \nu} F^{k-1})
\end{eqnarray}
where $F \equiv F_{\mu \nu} F^{\mu \nu}$ is the Maxwell invariant, while $G_{\mu \nu}$ is the Einstein tensor. We seek spherically static solutions of the form
\begin{equation}
ds^2 = -f(r) dt^2 + f(r)^{-1} dr^2 + r^2 d \phi^2
\end{equation}
where the metric function is found to be \cite{BH1,BH2}
\begin{equation}
f(r) = -M - \frac{1}{2}\pi  \frac{ (2k-1)^2}{ (k-1)} Q^{2k} r^{\frac{2 (k-1)}{2k-1}}
\end{equation}
where $M,Q$ are the mass and the electric charge of the black hole respectively. In the following we will study the propagation of a probe canonical massless scalar field into a given gravitational background with $k=3/4$. For this
value the metric function becomes
\begin{equation}
f(r) = -M + \frac{q}{r}
\end{equation}
where $q$ is defined to be $q=\pi Q^{3/2}/2$ , and the single event horizon is given by $r_H=q/M$. Then the metric function takes the form
\begin{equation}
f(r) = -M \left(1-\frac{r_H}{r}\right)
\end{equation}
We use natural units such that $c = 8 G = \hbar = 1$ and metric signature $(-, +, +)$.

\subsection{The scalar wave equation}

Next we consider in the above gravitational background a probe minimally coupled massless scalar field with equation of motion
\begin{equation}
\frac{1}{\sqrt{-g}} \partial_\mu (\sqrt{-g} g^{\mu \nu} \partial_\nu) \Phi = 0
\end{equation}
Using the standard ansatz
\begin{equation}\label{separable}
\Phi(t,r,\phi) = e^{i \omega t} R(r) e^{i m \phi}
\end{equation}
where $\omega$ is the frequency and $m$ is the quantum number of angular momentum, we obtain an ordinary differential equation for the radial part
\begin{equation}
R'' + \left( \frac{1}{r} + \frac{f'}{f} \right) R' + \left( \frac{\omega^2}{f^2} - \frac{m^2}{r^2 f} \right) R = 0
\end{equation}
where the prime denotes differentiation with respect to radial distance $r$. To see the effective potential barrier that the scalar field 
feels we define new variables as follows
\begin{eqnarray}
R & = & \frac{\psi}{\sqrt{r}} \\
x & = & \int \frac{dr}{f(r)}
\end{eqnarray}
where we are using the so-called tortoise coordinate $x$, given approximately by
\begin{equation}
x \simeq -\frac{r_H}{M} \ln(r-r_H)
\end{equation}
close to the horizon, and we recast the equation for the radial part into a Schr{\"o}dinger-like equation of the form
\begin{equation}
\frac{d^2 \psi}{dx^2} + (\omega^2 - V(x)) \psi = 0
\end{equation}
Therefore we obtain for the effective potential barrier the expression
\begin{equation}
V(r) = f(r) \: \left( \frac{m^2}{r^2}+\frac{f'(r)}{2 r} - \frac{f(r)}{4 r^2} \right)
\end{equation}
Since the effective potential barrier vanishes at the horizon, close to the horizon $\omega^2 \gg V(x)$, and the solution for the
Schr{\"o}dinger-like equation is given by
\begin{equation}
\psi(x) = A_- e^{-i \omega x} + A_+ e^{i \omega x}
\end{equation}
Requiring purely ingoing solution \cite{kanti1, Fernando:2004ay, chinos} we set $A_-=0$ in the following.
The effective potential as a function of the the tortoise coordinate is shown in Fig. 1 below. We have considered three different parameter pairs as follows ($M=1$, $Q=0.33$), ($M=1.1$, $Q=0.35$) and ($M=1.2$, $Q=0.37$). In all three cases the horizon takes the same value $r_H=0.3$. We can see that the effective potential has a nice gaussian--like shape, as expected for asymptotically flat spacetimes. In addition, the location of the maximum as well as the the height of the potential depend on the values of the mass and the charge of the black hole.
\begin{figure}[ht!]
\centering
\includegraphics[width=\linewidth]{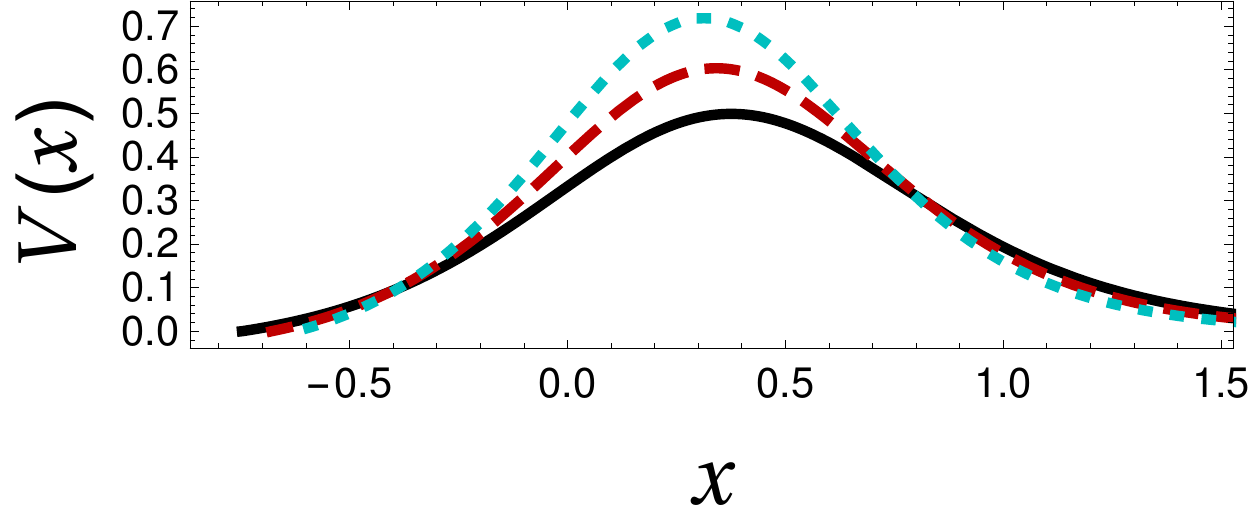}
\caption{\label{fig:1} 
The effective potential as a function of the tortoise coordinate $V(x)$ assuming $m = 0$ for three different cases: i) $M = 1.0$ and $Q = 0.33$ (solid black line), ii) $M = 1.1$ and $Q = 0.35$ (short dashed red line) and iii) $M=1.2$ and $Q = 0.37$ (long dashed blue line).
}
\label{fig:1} 	
\end{figure}

\section{Absorption cross section in the low energy regime} \label{Sol}

In this section we solve the radial differential equation analytically. Since exact solutions hardly exist, we find an approximate solution valid
in the low energy regime following a standard procedure described in \cite{Harmark:2007jy}. It consists of solving the radial equation in the far-field
and in the near-horizon regions, and then matching the solutions in the intermediate region.

\subsection{Solution in the far-zone regime}

When $r \gg r_H$ the metric function $f(r) \rightarrow -M$, and thus the radial equation takes the form
\begin{equation}
R'' + \frac{1}{r} R' + \left( \frac{\omega^2}{M^2} + \frac{m^2}{M r^2} \right) R = 0
\end{equation}
which can be recast into the Bessel equation of order $n=i \nu$ \cite{handbook}, with $\nu=|m|/\sqrt{M}$, and therefore the general solution is given by
\begin{equation}
R_{FF}(r) = B_+ J_{i \nu}\left(\frac{\omega r}{M}\right) + B_- Y_{i \nu}\left(\frac{\omega r}{M}\right)
\end{equation}
where $B_+, B_-$ are two arbitrary coefficients. In the limit $\omega r/M \rightarrow \infty$ the Bessel functions
behave asymptotically as plane waves \cite{handbook}, and thus the solution in the far-field region takes the form
\begin{align}
\begin{split}
R_{FF}(r) \simeq & \ \frac{B_+ - i B_-}{\sqrt{2 \pi \omega r/M}} e^{i (\omega r/M - \frac{\pi}{4})} e^{\pi \nu/2}  \  + 
\\
&
\
\frac{B_+ + i B_-}{\sqrt{2 \pi \omega r/M}} e^{-i (\omega r/M - \frac{\pi}{4})} e^{-\pi \nu/2}
\end{split}
\end{align}
and which leads to the following expression for the reflection coefficient
\begin{equation}
\mathcal{R} = \left | \frac{B_+ - i B_-}{B_+ + i B_-} e^{\pi \nu} \right |^2
\end{equation}
or defining the ratio $\tilde{B} \equiv B_+/B_-$, to be determined later on upon matching the solutions in the intermediate regime,
\begin{equation}
\mathcal{R} = \left | \frac{\tilde{B} - i}{\tilde{B} + i} e^{\pi \nu} \right |^2
\end{equation}
and finally we obtain the absorption cross section using the three-dimensional optical theorem relation \cite{optical,kanti1,Fernando:2004ay}
\begin{align}
\sigma_m (\omega) = \frac{1}{\omega} (1-\mathcal{R})=\frac{1}{\omega} \left(1-\left | \frac{\tilde{B} - i}{\tilde{B} + i} e^{\pi \nu} \right |^2 \right)
\end{align}

\subsection{Solution in the near-horizon regime}

We define a new dimensionless parameter as follows \cite{chinos}
\begin{equation}
z = 1- \frac{r_H}{r}
\end{equation}
that takes values in the range $0 < z < 1$. In the near-horizon regime $z \rightarrow 0$ the radial differential equation with respect to $z$ becomes
\begin{equation}
z (1-z) R_{zz} + (2-z) R_z + \left( \frac{A}{z} + \frac{B}{-1+z} \right) R = 0
\end{equation}
where the constants $A, B$ are given by
\begin{eqnarray}
A & = & \frac{\omega^2 r_H^2}{M^2} \\
B & = & - \left( \frac{\omega^2 r_H^2}{M^2} + \frac{m^2}{M} \right)
\end{eqnarray}
To get rid of the poles we set
\begin{equation}
R(z) = z^\alpha (1-z)^\beta F(z)
\end{equation}
where now the few function $F(z)$ satisfies the following differential equation
\begin{multline}
z (1-z) F_{zz} + [1+2 \alpha - (2+2 \alpha+2 \beta) z] F_z
\\
+ \left( \frac{\bar{A}}{z} + \frac{\bar{B}}{-1+z} - C \right) F = 0
\end{multline}
and the new constants are given by
\begin{eqnarray}
\bar{A} & = & A + \alpha^2 \\
\bar{B} & = & B - \beta^2 \\
C & = & (\alpha+\beta)^2+\alpha+\beta
\end{eqnarray}
Demanding that $\bar{A} = 0 = \bar{B}$ we obtain the
Gauss' hypergeometric equation
\begin{equation}
z (1-z) F_{zz} + [c-(1+a+b) z] F_z - ab F = 0
\end{equation}
and we determine the parameters $\alpha, \beta$ as follows
\begin{eqnarray}
\alpha & = & \pm i \frac{\omega r_H}{M}  \\
\beta & = & \pm i \sqrt{\frac{\omega^2 r_H^2}{M^2} + \frac{m^2}{M}} \label{beta}
\end{eqnarray}
Finally the three parameters of Gauss' equation are given by
\begin{eqnarray}
c & = & 1+2 \alpha \\
a & = &  \alpha + \beta + 1 \\
b & = & \alpha + \beta
\end{eqnarray}
Note that the parameters $a,b,c$ satisfy the condition $c-a-b=-2 \beta$.
Therefore the general solution for the radial part in the near-horizon region is given by
\begin{align}
\begin{split}
R_{NH}(z) = & \ z^\alpha (1-z)^\beta \bigl[ C_1 F(a,b;c;z) \ + 
\\
& \
C_2 z^{1-c} F(a-c+1,b-c+1;2-c;z) \bigl]
\end{split}
\end{align}
where $C_1,C_2$ are two arbitrary coefficients, and the hypergeometric function can be expanded in a Taylor series \cite{handbook}
as follows
\begin{equation}
F(a,b;c;z) = 1 + \frac{a b}{c} \:z + ...
\end{equation}
To recover the purely ingoing solution in the near-horizon regime we choose for $\alpha$ the minus sign and we set $C_2=0$, while for $\beta$ without loss
of generality we choose the plus sign. Therefore the solution becomes
\begin{equation}
R_{NH}(z) = D z^\alpha (1-z)^\beta F(a,b;c;z)
\end{equation}
where we have replaced $C_1$ by $D$.

\subsection{Matching of the solutions}

In this final step we stretch the solutions $R_{FF}(r)$ in the far-field region $r \gg r_H$ and $R_{NH}(z)$ in the near-horizon region $z \rightarrow 0$ to match them in the intermediate region. On the one hand, the $R_{FF}(r)$ expressed in terms of the Bessel functions in the limit $\omega r \ll 1$ becomes \cite{handbook}
\begin{equation}
R_{FF}(r) \simeq  \frac{B_+}{\Gamma(1+i \nu)} \left( \frac{\omega r}{2M} \right)^{i \nu} - \frac{B_- \Gamma(i \nu)}{\pi} \left( \frac{\omega r}{2M} \right)^{-i \nu}
\end{equation}
On the other hand, first we use the transformation formula \cite{handbook}
\begin{equation}
\begin{split}
F(a,b;c;z) = \ &\frac{\Gamma(c) \Gamma(c-a-b)}{\Gamma(c-a) \Gamma(c-b)}
\ \times
\\
&F(a,b;a+b-c+1;1-z) \ +
\\
 (1-z)^{c-a-b} &\frac{\Gamma(c) \Gamma(a+b-c)}{\Gamma(a) \Gamma(b)}
\ \times
\\
&F(c-a,c-b;c-a-b+1;1-z)
\end{split}
\end{equation}
and therefore the near-horizon solution as $z \rightarrow 1$ reads
\begin{align}
\begin{split}
R_{NH}(z \rightarrow 1) = & \  \frac{D (1-z)^\beta \Gamma(1+2 \alpha) \Gamma(-2 \beta)}{\Gamma(\alpha-\beta) \Gamma(1+\alpha-\beta)} \ + 
\\
& \  \frac{D (1-z)^{-\beta} \Gamma(1+2 \alpha) \Gamma(2 \beta)}{\Gamma(1+\alpha+\beta) \Gamma(\alpha+\beta)}
\end{split}
\end{align}
Since $1-z=r_H/r$ and in the low energy regime $\beta \simeq i \nu$, matching the two solutions we obtain the ratio $\frac{B_+}{B_-} = \tilde{B} $
\begin{equation}
\tilde{B} = \left[ \frac{2M}{\omega r_H} \right]^{2 i \nu}
 \frac{\Gamma(2 \beta) \Gamma(\alpha-\beta) \Gamma(1+\alpha-\beta)  \nu \Gamma(i \nu)^2 }{i \pi \Gamma(-2 \beta) \Gamma(1+\alpha+\beta) \Gamma(\alpha+\beta)} 
\end{equation}
This is the main result of the present article. This formula allows us to compute the reflection coefficient using eq. (21) and the greybody factor using eq. (22).
%
%
%
\begin{figure}[ht!]
\centering
\includegraphics[width=\linewidth]{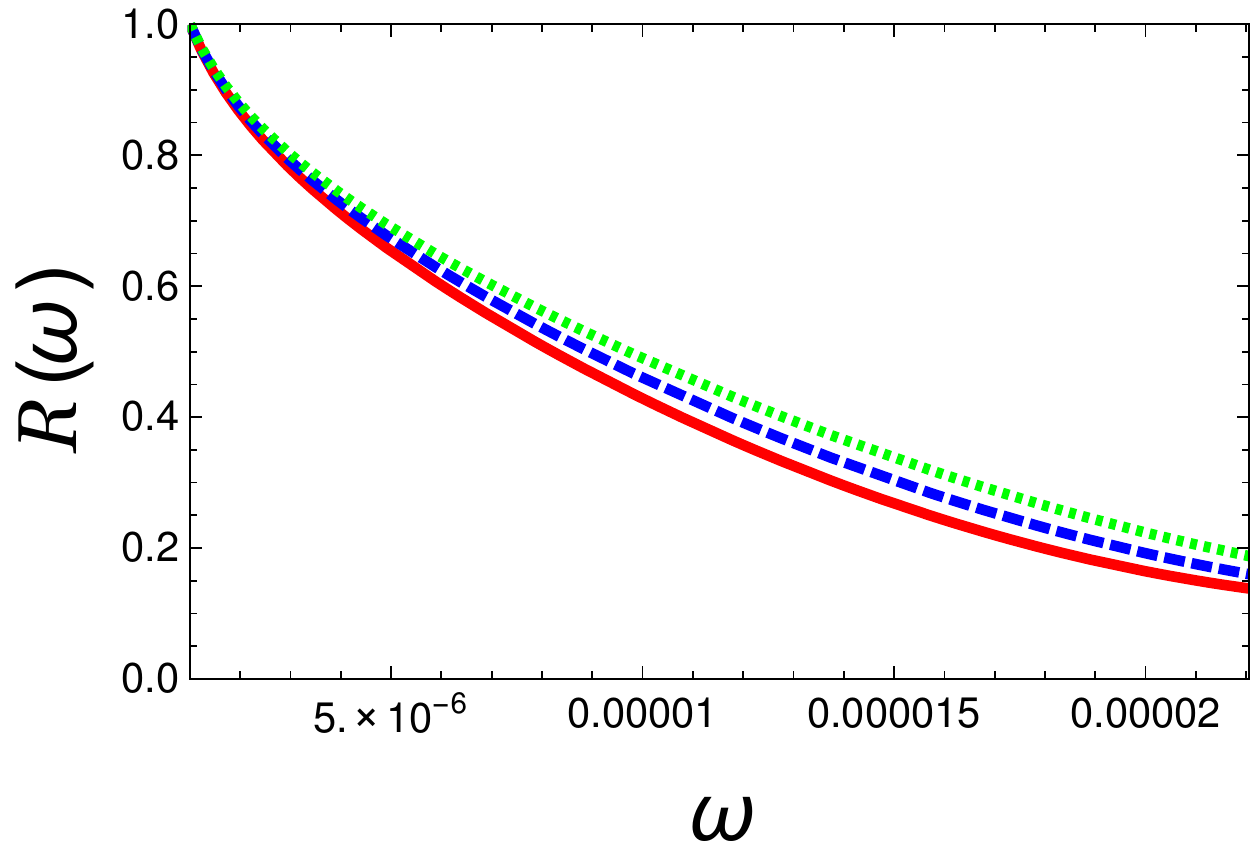}
\caption{\label{fig:2} 
The reflection coefficient $\mathcal{R}$ as a function of the frequency $\omega$ assuming $m = 1$ for three different cases as follows (from bottom to top): i) $M = 140$ and $Q = 9.22 \times 10^4$ (solid red line), ii) $M = 145$ and $Q = 9.02 \times 10^4$ (dashed blue line) and iii) $M=150$ and $Q = 8.82 \times 10^4$ (dotted green line). 
}
\end{figure}

\subsection{Brief discussion of the results}

\begin{figure}[ht!]
\centering
\includegraphics[width=\linewidth]{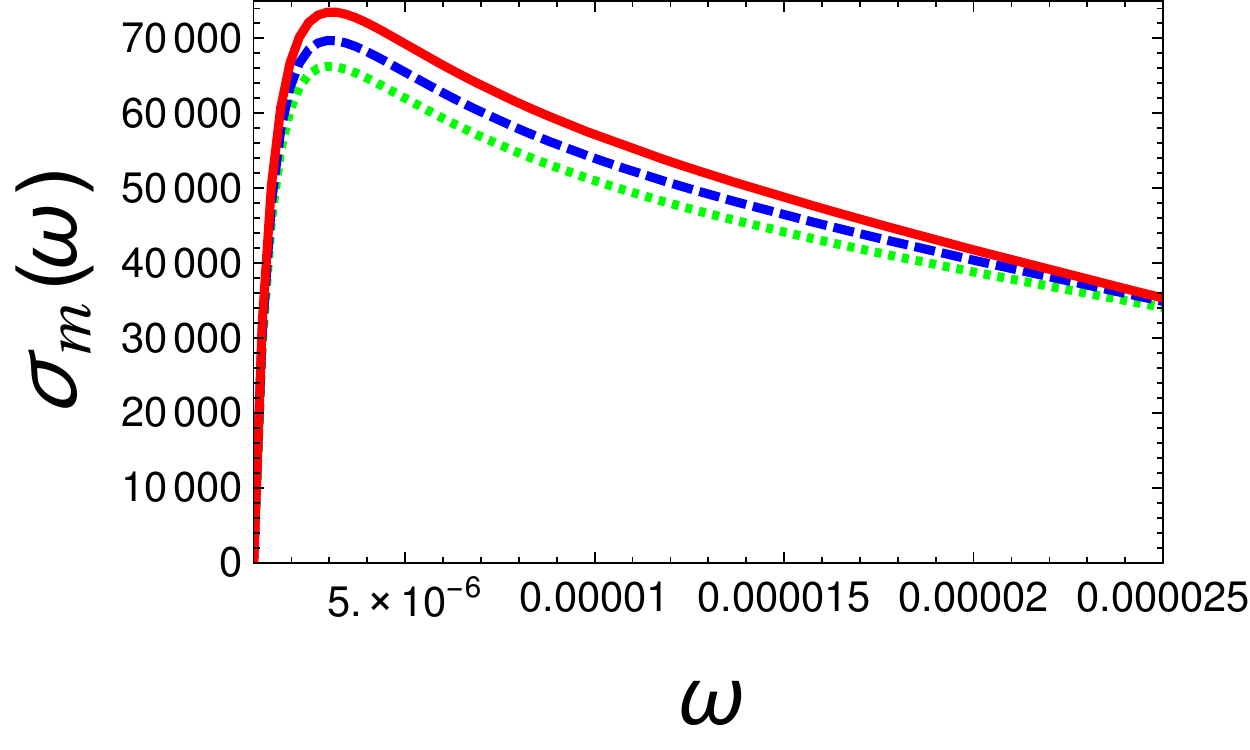}
\caption{\label{fig:3}
The greybody factor $\sigma_{m}$ as a function of the frequency $\omega$ assuming $m = 1$ for three different cases as follows:
(from top to bottom) i) $M = 140$ and $Q = 9.22 \times 10^4$ (solid red line), ii) $M = 145$ and $Q = 9.02 \times 10^4$ (dashed blue line) and iii) $M=150$ and $Q = 8.82 \times 10^4$ (dotted green line). 
}
\end{figure}

Next we plot the coefficient of reflection $\mathcal{R}$ and the greybody factor $\sigma_m$ versus frequency $\omega$ in Figures \ref{fig:2} and \ref{fig:3} respectively. First $\mathcal{R}$ versus $\omega$ for $m=1$ and three different ($M,Q$) pairs is shown in Fig. \ref{fig:2}. The black hole mass is chosen such that $M \gg 1$ where the semi-classical approximation is valid. The reflection coefficient starts from 1 and decreases monotonically to zero as it should. From general principles we know that eventually it tends to zero. Here, however, we are not allow to plot it up to higher values of the frequency since the formula we have obtained can be trusted only in the low energy regime where the term $m^2/M$ dominates over $(\omega r_H/M)^2$. Furthermore, we observe that increasing the mass $M$ and lowering the charge $Q$ of the black hole the curves are shifted upwards.
%
%
%

The greybody factor $\sigma_m$ as a function of $\omega$ for $m=1$ and for the same 3 different ($M,Q$) pairs is shown in Fig. \ref{fig:3}. We observe that the cross section starts from zero (as it is expected for any non-vanishing quantum number of angular momentum) and increases with $\omega$ until it reaches a maximum, and then it
decreases monotonically tending to zero as it is expected from general principles. However, given that our expression can be trusted only in the low energy regime, we are not allowed to plot it versus the energy up to higher values, and like Fig. \ref{fig:2} we have restricted ourselves to the range in which $m^2/M > (\omega r_H/M)^2$. In addition, increasing the mass and lowering the charge the curves are shifted downwards.

\section{Conclusions}
\label{Concl}

In this article we have studied the propagation of a probe minimally coupled massless scalar field in a three-dimensional Einstein-power-Maxwell charged black hole spacetime. We have considered spherically symmetric static backgrounds that are characterized by two free parameters related to the mass and the charge of the black hole. Applying standard techniques first we recast the radial part of the scalar wave equation into a Schr{\"o}dinger-like equation, and we read off the effective potential barrier. Then we solve the radial differential equation in the far-zone regime, in the near horizon regime, and we finally stretch the solutions to match them in the intermediate regime. Finally, the dependence of the reflection coefficient as well as of the greybody factor on the parameters of the theory is discussed.


\begin{acknowledgments}
The work of A.R. was supported by the CONICYT-PCHA/Doctorado Nacional/2015-21151658.
G.P. thanks the Funda\c c\~ao para a Ci\^encia e Tecnologia (FCT), Portugal, for the financial support to the Center for Astrophysics and Gravitation-CENTRA, Instituto Superior T\'ecnico, Universidade de Lisboa, through the Grant No. UID/FIS/00099/2013.
\end{acknowledgments}



\begin{thebibliography}{99}
\bibitem{hawking1} S.~W.~Hawking,
  Nature {\bf 248} (1974) 30.
  
\bibitem{hawking2} S.~W.~Hawking,
  Commun.\ Math.\ Phys.\  {\bf 43} (1975) 199
   Erratum: [Commun.\ Math.\ Phys.\  {\bf 46} (1976) 206].
   
\bibitem{kanti1} P.~Kanti and J.~March-Russell,
  Phys.\ Rev.\ D {\bf 66} (2002) 024023
  [hep-ph/0203223].
\bibitem{col1}  C. Doran,  A. Lasenby,  S. Dolan,  and  I.  Hinder,  Phys. Rev. D {\bf 71}, 124020 (2005).

\bibitem{col2} S. Dolan, C. Doran, and A. Lasenby, Phys. Rev. D {\bf 74}, 064005 (2006).

\bibitem{col3} L. C. B. Crispino, E. S. Oliveira, A. Higuchi, and G. E. A. Matsas, Phys. Rev. D {\bf 75}, 104012 (2007).

\bibitem{col4} S.  R.  Dolan,  Classical  Quantum  Gravity {\bf 25}, 235002 (2008).

\bibitem{col5} L. C. B. Crispino, S. R. Dolan, and E. S. Oliveira,
Phys. Rev. Lett. {\bf 102}, 231103 (2009).

\bibitem{col6} L. C. B. Crispino, S. R. Dolan, and E. S. Oliveira,
Phys. Rev. D {\bf 79}, 064022 (2009).

\bibitem{col7} L. B. Crispino, A. Higuchi, and E. S. Oliveira,
Phys. Rev. D {\bf 80}, 104026 (2009).

\bibitem{3D1} D.~Birmingham, I.~Sachs and S.~Sen,
  Phys.\ Lett.\ B {\bf 413} (1997) 281
  [hep-th/9707188].
  
\bibitem{3D2} Y.~S.~Myung,
  Mod.\ Phys.\ Lett.\ A {\bf 18} (2003) 617
  [hep-th/0201176].
  
\bibitem{Fernando:2004ay} S.~Fernando,
  Gen.\ Rel.\ Grav.\  {\bf 37} (2005) 461,
  [hep-th/0407163].
  
\bibitem{chinos} Y.~Liu and J.~L.~Jing,
  Chin.\ Phys.\ Lett.\  {\bf 29} (2012) 010402.
  
\bibitem{coupling} L.~C.~B.~Crispino, A.~Higuchi, E.~S.~Oliveira and J.~V.~Rocha,
  Phys.\ Rev.\ D {\bf 87} (2013) 104034
  [arXiv:1304.0467 [gr-qc]].
  
\bibitem{kanti2} P.~Kanti, T.~Pappas and N.~Pappas,
  Phys.\ Rev.\ D {\bf 90} (2014) no.12,  124077
  [arXiv:1409.8664 [hep-th]].
  
\bibitem{kanti3} T.~Pappas, P.~Kanti and N.~Pappas,
  Phys.\ Rev.\ D {\bf 94} (2016) no.2,  024035
  [arXiv:1604.08617 [hep-th]].
  
\bibitem{Panotopoulos:2016wuu} G.~Panotopoulos and \'A.~Rinc\'on,
  Phys.\ Lett.\ B {\bf 772}, 523 (2017)
[arXiv:1611.06233 [hep-th]].
 
\bibitem{Panotopoulos:2017yoe} G.~Panotopoulos and \'A.~Rinc\'on,
  Phys.\ Rev.\ D {\bf 96}, no. 2, 025009 (2017)
[arXiv:1706.07455 [hep-th]].
%
\bibitem{Ahmed:2016lou} J.~Ahmed and K.~Saifullah,
  arXiv:1610.06104 [gr-qc].
  
\bibitem{CS}
A.~Achucarro and P.~K. Townsend, ``{A Chern-Simons Action for Three-Dimensional
  anti-De Sitter Supergravity Theories},'' {\em Phys. Lett.}, vol.~B180, p.~89,
  1986.

\bibitem{Witten:1988hc}
E.~Witten, ``{(2+1)-Dimensional Gravity as an Exactly Soluble System},'' {\em
  Nucl. Phys.}, vol.~B311, p.~46, 1988.

\bibitem{Witten:2007kt}
E.~Witten, ``{Three-Dimensional Gravity Revisited},'' 2007.

\bibitem{review}
S.~Carlip, ``{The (2+1)-Dimensional black hole},'' {\em Class. Quant. Grav.},
  vol.~12, pp.~2853--2880, 1995.

\bibitem{BI} M.~Born and L.~Infeld,
Proc.\ Roy.\ Soc.\ Lond.\ A {\bf 144} (1934) 425.

\bibitem{ST1} M. B. Green, J. H. Schwarz and E. Witten, \textit{Superstring Theory, Vol. 1 \& 2},
Cambridge Monographs on Mathematical Physics.

\bibitem{ST2} J. Polchinski, \textit{String Theory, Vol. 1 \& 2}, Cambridge Monographs on Mathematical Physics.

\bibitem{Dbranes1} C. V. Johnson, \textit{D-Branes}, Cambridge Monographs on Mathematical Physics.

\bibitem{Dbranes2} B. Zwiebach, \textit{A First Course in String Theory }, Cambridge University Press.

\bibitem{Xu:2014uka} W.~Xu and D.~C.~Zou,
  Gen.\ Rel.\ Grav.\  {\bf 49}, no. 6, 73 (2017)
  [arXiv:1408.1998 [hep-th]].
  
\bibitem{Mazharimousavi:2011nd} S.~H.~Mazharimousavi, O.~Gurtug, M.~Halilsoy and O.~Unver,
  Phys.\ Rev.\ D {\bf 84}, 124021 (2011)
  [arXiv:1103.5646 [gr-qc]].
  
\bibitem{Rincon:2017goj} \'A.~Rinc\'on, E.~Contreras, P.~Bargue\~no, B.~Koch, G.~Panotopoulos and A.~Hern\'andez-Arboleda,
  Eur.\ Phys.\ J.\ C {\bf 77}, no. 7, 494 (2017)
  [arXiv:1704.04845 [hep-th]].

\bibitem{Panotopoulos:2017hns} 
  G.~Panotopoulos and \'A.~Rinc\'on,
  Int.\ J.\ Mod.\ Phys.\ D {\bf }
  doi:10.1142/S0218271818500347
  [arXiv:1711.04146 [hep-th]].

\bibitem{Rincon:2018sgd} 
  \'A.~Rinc\'on and G.~Panotopoulos,
  Phys.\ Rev.\ D {\bf 97}, no. 2, 024027 (2018)
  doi:10.1103/PhysRevD.97.024027
  [arXiv:1801.03248 [hep-th]].

\bibitem{BH1} O.~Gurtug, S.~H.~Mazharimousavi and M.~Halilsoy,
  Phys.\ Rev.\ D {\bf 85} (2012) 104004
[arXiv:1010.2340 [gr-qc]].

\bibitem{BH2} M.~Hassaine and C.~Martinez,
  Class.\ Quant.\ Grav.\  {\bf 25} (2008) 195023
[arXiv:0803.2946 [hep-th]].

\bibitem{Harmark:2007jy} T.~Harmark, J.~Natario and R.~Schiappa,
  Adv.\ Theor.\ Math.\ Phys.\  {\bf 14} (2010) no.3,  727
[arXiv:0708.0017 [hep-th]].

\bibitem{handbook} M. Abramowitz and I. Stegun, \textit{Handbook of Mathematical Functions} (Academic, New York, 1966).

\bibitem{optical} S.~S.~Gubser, I.~R.~Klebanov and A.~A.~Tseytlin,
  Nucl.\ Phys.\ B {\bf 499} (1997) 217
[hep-th/9703040].
\end{thebibliography}
\end{document}